\documentclass[12pt]{iopart}

\begin{document}

\title{Determination of the Shear Viscosity Relaxation Time at Weak and
Strong Coupling}

\author{G.\ S.\ Denicol${}^{1}$, J.\ Noronha${}^{2}$, H.\ Niemi${}^{3}$, and
D.\ H.\ Rischke${}^{1,3}$ }

\address{$^{1}$Institut f\"ur Theoretische Physik, Johann Wolfgang
Goethe-Universit\"at, Max-von-Laue-Str.\ 1, D-60438 Frankfurt am Main,
Germany} 
\address{$^{2}$ Instituto de F\'{i}sica, Universidade de S\~{a}o Paulo, C.P.
66318, 05315-970 S\~{a}o Paulo, SP, Brazil} 
\address{$^{3}$Frankfurt Institute for Advanced Studies, Ruth-Moufang-Str.\ 1,
D-60438 Frankfurt am Main, Germany}

\begin{abstract}
We investigate the microscopic origin of the relaxation time coefficient in
relativistic fluid dynamics. We show that the extraction of the shear
viscosity relaxation time via the gradient expansion is ambiguous and in
general fails to give the correct result. The correct value for the shear
viscosity relaxation time is extracted from the slowest non-hydrodynamic
pole of the corresponding retarded Green's function, if such a pole is
purely imaginary. According to the AdS/CFT correspondence, in
strongly-coupled $\mathcal{N}=4$ SYM the non-hydrodynamic poles of the shear
stress tensor nearest to the origin have a nonzero real part, which implies
that the transient fluid-dynamical equations for this gauge theory are not
equivalent to the well-known Israel-Stewart equations.
\end{abstract}

The derivation of a consistent theory of \textit{relativistic} fluid
dynamics has been a challenge for some time. The difficulty resides in the
parabolic nature of Navier-Stokes theory, which allows signals that
propagate with infinite speed. While in nonrelativistic theories this
feature can be overlooked, in relativistic systems causality is intimately
related to stability \cite{stability} and this feature leads to unstable
equations of motion. A causal and stable theory of fluid dynamics can be
obtained by taking into account the \textit{transient} dynamics of the
dissipative currents \cite{IS}, i.e., by considering the characteristic
times within which fluid-dynamical dissipative currents, such as the shear
stress tensor $\pi ^{\mu \nu }$, relax towards their asymptotic
Navier-Stokes solution.

In this contribution, we investigate the physical origin of such
characteristic time scales for transient fluid behavior, called the
relaxation time coefficients \cite{IS}. We discuss two distinct formulations
of relativistic fluid dynamics: The first, derived in Ref.~\cite{BRSSS},
extends Navier-Stokes theory via the gradient expansion, for the specific
case of conformal fluids, and introduces the relaxation time as a type of
regulator to obtain stable fluid-dynamical equations. This procedure is
implemented in such a way that the regularized theory is equivalent to the
gradient expansion for low frequencies and large wavelengths. The second
approach, originally presented in Ref.~\cite{poles}, derives fluid dynamics
from the underlying microscopic theory taking into account the slowest
non-hydrodynamic mode for each dissipative current. The theory obtained in
this case is not only equivalent to the gradient expansion for low
frequencies and large wavelengths, it is also able to describe the transient
dynamics of the dissipative currents. In this contribution, we consider only
the case of the shear stress tensor.

In Ref.~\cite{BRSSS}, a relativistic Burnett equation is derived for
conformal fluids. The basic idea of this derivation is to extend
Navier-Stokes theory by including all possible terms of second order in
gradients of temperature $T$ and four-velocity $u^{\mu }$ that are
symmetric, transverse, traceless, and that transform homogeneously under
Weyl transformations \cite{BRSSS},%
\begin{eqnarray*}
\mathcal{O}_{1}^{\mu \nu }= R^{\left\langle \mu \nu \right\rangle
}-2\left( \nabla ^{\left\langle \mu \right. }\nabla ^{\left. \nu
\right\rangle }\ln T-\nabla ^{\left\langle \mu \right. }\ln T\nabla ^{\left.
\nu \right\rangle }\ln T\right) , \\
\mathcal{O}_{2}^{\mu \nu}=R^{\left\langle \mu \nu \right\rangle }-2u_{\alpha }R^{\alpha \left\langle
\mu \nu \right\rangle \beta }u_{\beta }, \\
\mathcal{O}_{3}^{\mu \nu } =\sigma _{\lambda }^{\langle \mu }\sigma ^{\nu
\rangle \lambda },~~\mathcal{O}_{4}^{\mu \nu }=\sigma _{\lambda }^{\langle
\mu }\omega ^{\nu \rangle \lambda },~~\mathcal{O}_{5}^{\mu \nu }=\omega
_{~\lambda }^{\langle \mu }\omega ^{\nu \rangle \lambda },
\end{eqnarray*}%
with $\sigma ^{\mu \nu }$ being the shear tensor, $\omega ^{\mu \nu }$ the
vorticity tensor, $R^{\mu \nu }$ the Ricci tensor, and $R^{\mu \nu \alpha
\beta }$ the Riemann tensor. We adopt the notation $A^{\left\langle \mu \nu
\right\rangle }=\Delta _{\alpha \beta }^{\mu \nu }A^{\alpha \beta }$ with $%
\Delta _{\alpha \beta }^{\mu \nu }=\left( \Delta _{\alpha }^{\mu }\Delta
_{\beta }^{\nu }+\Delta _{\beta }^{\mu }\Delta _{\alpha }^{\nu }-2/3\Delta
^{\mu \nu }\Delta _{\alpha \beta }\right) /2$ being the traceless and
symmetric projection operator and $\Delta ^{\mu \nu }=g^{\mu \nu }+u^{\mu
}u^{\nu }$ being the projection operator orthogonal to $u^{\mu }$. Then, the
shear stress tensor, up to second order in gradients, is assumed to be%
\begin{equation}
\pi ^{\mu \nu }=-2\eta \sigma ^{\mu \nu }+\sum_{i=1}^{5}\lambda _{i}\mathcal{%
O}_{i}^{\mu \nu }.  \label{2}
\end{equation}

As is well-known, the relativistic and nonrelativistic Burnett equations are
unstable and have no practical use. In order to render the theory stable, a
relaxation time coefficient for the shear stress tensor must be introduced.
In Ref.~\cite{BRSSS} this was done in the following way. Using the
conservation laws of energy and momentum it is possible to show that, up to
third order in gradients,%
\begin{equation}
\dot{\sigma}^{\left\langle \mu \nu \right\rangle }+\frac{1}{3}\sigma ^{\mu
\nu }\theta =\mathcal{O}_{1}^{\mu \nu }-\mathcal{O}_{2}^{\mu \nu }-\frac{1}{2%
}\mathcal{O}_{3}^{\mu \nu }+2\mathcal{O}_{5}^{\mu \nu }+\mathcal{O}\left(
\nabla ^{3}\right) .  \label{trick}
\end{equation}%
This equation is then used to replace the term $\mathcal{O}_{1}^{\mu \nu }$
in Eq.~(1), i.e.,%
\begin{eqnarray}
\pi ^{\mu \nu } &=&-2\eta \sigma ^{\mu \nu }+\lambda _{1}\left( \dot{\sigma}%
^{\left\langle \mu \nu \right\rangle }+\frac{1}{3}\sigma ^{\mu \nu }\theta
\right) +\left( \lambda _{2}+\lambda _{1}\right) \mathcal{O}_{2}^{\mu \nu } 
\nonumber \\
&&+\left( \lambda _{3}+\frac{1}{2}\lambda _{1}\right) \mathcal{O}_{3}^{\mu
\nu }+\lambda _{4}\mathcal{O}_{4}^{\mu \nu }+\left( \lambda _{5}-2\lambda
_{1}\right) \mathcal{O}_{5}^{\mu \nu }.  \label{3}
\end{eqnarray}%
The last step of the derivation is to construct a relaxation equation for $%
\pi ^{\mu \nu }$ by substituting the first-order solution, $\sigma ^{\mu \nu
}=-\left( 1/2\eta \right) \pi ^{\mu \nu }$, in all second-order terms%
\begin{eqnarray}
\tau _{\pi }\dot{\pi}^{\left\langle \mu \nu \right\rangle }+\pi ^{\mu \nu }
&=&-2\eta \sigma ^{\mu \nu }-\frac{4}{3}\tau _{\pi }\pi ^{\mu \nu }\theta
+\left( \lambda _{2}+\lambda _{1}\right) \mathcal{O}_{2}^{\mu \nu } 
\nonumber \\
&&+\left( \lambda _{3}+\frac{1}{2}\lambda _{1}\right) \tilde{\mathcal{O}}%
_{3}^{\mu \nu }+\lambda _{4}\tilde{\mathcal{O}}_{4}^{\mu \nu }+\left(
\lambda _{5}+2\lambda _{1}\right) \mathcal{O}_{5}^{\mu \nu },  \label{BRSSS}
\end{eqnarray}%
where $\tilde{\mathcal{O}}_{3,4}^{\mu \nu }$ corresponds to $\mathcal{O}%
_{3,4}^{\mu \nu }$ with the substitution $\sigma ^{\mu \nu }\rightarrow -\pi
^{\mu \nu }/(2\eta )$ and the coefficient $\tau _{\pi }$ was identified as $%
\lambda _{1}/\left( 2\eta \right) $.

The coefficients $\lambda _{1,2}$ can be calculated using linear response
theory via metric perturbations. In strongly-coupled $\mathcal{N}=4$ SYM
theory, the coefficients $\lambda _{1}$ and $\eta $ were obtained from
derivatives of the retarded Green's function at the origin\footnote{%
The associated retarded Green's function in this case was computed using the
AdS/CFT correspondence \cite{review}.} and the relaxation time was computed
using this prescription, i.e., $\tau _{\pi }=\lambda _{1}/\left( 2\eta
\right) =\left( 2-\ln 2\right) /\left( 2\pi T\right) $.

Note, however, that this prescription to obtain the shear viscosity
relaxation time is ambiguous and, consequently, all the other transport
coefficients of Eq.~(\ref{BRSSS}) are also ill-defined. To the best of the
authors' knowledge, the substitution of the second-order term $\mathcal{O}%
_{1}^{\mu \nu }$ using Eq.~(\ref{trick}) is a choice. For example, if
instead of $\mathcal{O}_{1}^{\mu \nu }$, the term $\mathcal{O}_{2}^{\mu \nu }
$ is replaced using Eq.~(\ref{trick}), the relaxation time would be
identified as $\tau _{\pi }=\lambda _{2}/\left( 2\eta \right) $. Naturally,
there are infinitely many ways to make such a replacement and, in this
sense, infinitely different values for the shear viscosity relaxation time
could have been obtained in this approach. We remark that any choice would
satisfy the conditions set by the authors of Ref.~\cite{BRSSS} that the
regularized theory in Eq.~(\ref{BRSSS}) should be equivalent to the gradient
expansion at low frequencies and large wavelengths.

The reason for this ambiguity can be easily understood. Different choices of
replacements would describe different transient dynamics, described by their
respective relaxation time coefficients, but would invariably lead to the
same asymptotic solution, i.e., the Burnett equation in Eq.~(\ref{2}).
Naturally, it is always possible to determine the asymptotic dynamics of a
system once the transient dynamics is known, but the inverse is impossible.
Since the starting point of Ref.~\cite{BRSSS} is the asymptotic theory
itself, i.e., the gradient expansion. The actual transient dynamics with the
correct relaxation time cannot be uniquely extracted from this approach.

Recently, a new method to derive fluid dynamics was introduced in Ref.~\cite%
{poles}. In this approach, transient fluid dynamics is derived directly from
an underlying microscopic theory. Furthermore, this work derives a condition
for the retarded Green's function under which the linearized equation of
motion of a dissipative current can be reduced to a relaxation-type
equation. The equations of motion obtained were shown to have the gradient
expansion as its asymptotic solution, i.e., they are equivalent to the
Navier-Stokes and Burnett theories at low frequency and large wavelength.
However, these equations of motion are not only capable to describe the
asymptotic regime but they also correctly describe the dynamics of the shear
stress tensor at time scales of the order of the relaxation time, which was
shown to be related to the pole of the retarded Green's function closest to
the origin. This clears up a few misconceptions in the field such as whether
the relaxation time is of microscopic origin or of fluid-dynamic origin and
whether a truncated Taylor expansion around the origin can be used to
correctly obtain the relaxation time of dissipative fluid dynamics.

This approach was applied in Ref.~\cite{poles} to study dilute gases where
the Boltzmann equation can be used to describe the underlying microscopic
theory. In this work, microscopic formulas for the shear viscosity
coefficient, $\eta $, and relaxation time, $\tau _{\pi }$, were derived
using linear response theory. It was demonstrated that the retarded Green's
function associated with the shear stress tensor has infinitely many simple
poles along the imaginary axis making it possible to reduce the equation of
motion for the shear stress tensor to a relaxation-type equation. The
formulas for $\eta $ and $\tau _{\pi }$ were shown to be consistent with
those previously derived by Israel and Stewart \cite{IS,dkr} via the
14-moment approximation. This puts relativistic viscous hydrodynamics on a
solid ground.

For strongly-coupled conformal plasmas defined via the AdS/CFT
correspondence the situation was shown to be more subtle \cite{AdS}. For a
relaxation equation to exist, i.e., an equation of motion with only one time
derivative, it is mandatory for the pole of the retarded Green's function
nearest to the origin to be purely imaginary. In this case, the system obeys
a simple relaxation equation at long times, as it is the case with the
Boltzmann equation \cite{poles}. This, however, does not occur in strongly
coupled $\mathcal{N}=4$ SYM theory: in this theory the poles of the retarded
Green's function nearest to the origin have real parts and are symmetric
around the imaginary axis \cite{review}. Thus, both poles contribute equally to the long-time dynamics. Since the poles have a real part,
the system not only relaxes to its Navier-Stokes solution, but oscillates
around it as well. Then, the linearized equation of motion satisfied by the
shear stress tensor, at long times, should be of the form:%
\begin{equation}
\chi _{2}\tau _{\pi }\ddot{\pi}^{\left\langle \mu \nu \right\rangle }+\chi
_{1}\dot{\pi}^{\left\langle \mu \nu \right\rangle }+\pi ^{\mu \nu }=-2\eta
\sigma ^{\mu \nu }+\ldots .
\end{equation}%
Note that this equation was not assumed: It was directly derived from the
retarded Green's function of the microscopic theory and it describes the
long-time dynamics of the system. It should also be remarked that this
equation has the Navier-Stokes and Burnett theories as its asymptotic
solution. The coefficients of the equation above in the case of strongly
coupled $\mathcal{N}=4$ SYM theory were computed in Ref.~\cite{AdS} and
shown to be%
\begin{equation}
\chi _{1}=1.27/(4\pi T),~\chi _{2}=0.93/(4\pi T)^{2},~\eta /s=1/\left( 4\pi
\right) ,
\end{equation}%
where $T$ is the temperature and $s$ is the entropy density.

The authors thank the Helmholtz International Center for FAIR within the
framework of the LOEWE program for support. H.N. was supported by the
ExtreMe Matter Institute EMMI.

\end{document}